# Low-Cost Wideband Tilted Beam Antenna for Millimeter-wave Vehicle Applications

Jiawang Li

*Abstract*—To facilitate vehicle coverage for millimeter-wave applications, this communication presents a low-cost, wideband tilted-beam antenna. A novel design is proposed in which a slot antenna is both directly excited and electromagnetically coupled to a monopole array. This slot–monopole configuration is inherently robust against substrate losses, enabling low-cost fabrication while maintaining high realized gain and compact size. Furthermore, the slot-fed structure effectively excites multiple resonant modes within the monopole array, resulting in a significantly enhanced bandwidth. Experimental results demonstrate that the antenna achieves a -10-dB impedance bandwidth of over 76.5% (20–44.78 GHz) and a peak realized gain of 6.1 dBi.

## I. INTRODUCTION

With the rapid development of wireless communications, antennas play an increasingly important role in fulfilling tough performance requirements. In many applications, specific radiation patterns are needed, such as high-gain pattern from phased array antennas for 5G small base stations [1], monopole-like pattern for indoor WLAN [2], flat-top pattern for microwave power transmission [3] and tilted beam for smart transportation [4]. In particular, tilted beam antenna is also suitable for access points or sensors installed at the roadside communication device (RCD), as shown in Fig. 1(a). For this application, it is desirable to tilt the broadside beam by just over 45° to provide uniform coverage, since there are omnidirectional antennas on the roof of the car. For reasons of cost and simple integration, conventional access points or sensors often utilize planar broadside antennas, especially microstrip antennas. As shown in Fig. 1(b), a naive approach to provide the beam tilt is to mechanically tilt the entire device (hence the antenna) by a fixed angle $\theta$ during installation. However, this method not only increases the device's cross-sectional area and affects its aesthetics, it also complicates installation and accurate control of the tilt angle.

On the contrary, if the broadside pattern can be tilted as depicted in Fig. 1(c), the size of the installed device is reduced by $l \sin\theta - h$, where $l$ is the device length (along the tilted side) and $h$ is he thickness. The height reduction is considerable since typically $l \gg h$. Moreover, the installation process is simplified and the aesthetics is improved. However, for wideband operation, it is very challenging to design tilted beam with significant overlapping 3 dB beamwidth across the large frequency band. This problem is well illustrated by frequency scanning antennas (e.g., leaky wave antennas) and arrays that intentionally employ the variation in their electrical lengths across frequency to achieve scanning of the beam direction [5]. Similarly, even though phased arrays can provide a fixed beam tilt by its scanning capability, a wideband design is challenging [6] and a larger footprint is needed for a multi-element antenna array. Alternatively, there are different means to realize tilted beam using single-port antennas, which can be categorized into: 1) metamaterial-

Manuscript received xxxx.xx.xx. (Corresponding author: *Jiawang Li*)

Jiawang Li is with the Department of Electrical and Information Technology, Lund University, 22100 Lund, Sweden (e-mail: jiawang.li@eit.lth.se).

Color versions of one or more of the figures in this communication are available online at http://ieeexplore.ieee.org.

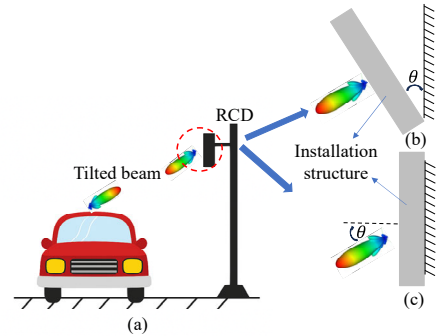

Fig. 1. Vehicle application of tilted beam: (a) home scenario, (b) mechanical beam tilt of broadside antenna, (c) tilted beam antenna.

based methods [7]-[15], 2) asymmetrical structure methods [16]-[17], and 3) multi-radiator methods [4], [18]. Metamaterial based methods achieve beam tilting by adding a metamaterial (often realized with multilayer substrate) above a radiating source (antenna element) to artificially modify the phase distribution of the wavefronts impinging on the metamaterial surface [7]-[15]. For example, by dimensioning $M \times (2N - 1)$ unit cells of square loops printed on a substrate placed above a patch antenna to provide different phase shifts along one dimension, the broadside beam of the antenna is tilted along that dimension [7]. The measured $9 \times 9$ case achieved 33° beam tilt, 16.4 dBi maximum gain and 10 dB impedance bandwidth of 5.2% [7]. Moreover, the beam tilt angle can be made reconfigurable by adding PIN diodes to electronically change the reflective phase of the unit cell, achieving a beam tilt of up to ±23° [14], [15]. The drawbacks of the metamaterial-based methods include the relatively large size of the metamaterial as well as limited impedance bandwidth and beam tilt angle (below 10% and 45°, respectively [7]-[15]).

In contrast, the asymmetric structure approaches [16]–[17] achieve beam tilting by introducing asymmetry into the radiating element. For instance, a 30°-tilted dielectric resonator antenna (TDRA) placed on a large ground plane achieves an average beam tilt of 30° and a gain exceeding 6 dBi over a bandwidth of 10.7% [16]. Similarly, by incorporating asymmetry into a broadband bowtie antenna, a 40° beam tilt and a peak gain of 5 dBi are achieved across a usable bandwidth of 32%, without employing any engineered materials [17]. While these designs operate based on intuitive principles, their performance is constrained either by limited bandwidth [16] or by relatively large physical dimensions [17].

The multi-radiator approach represents a novel beam-tilting concept, wherein a single feed simultaneously excites multiple radiating elements that exhibit distinct far-field patterns [4], [18]. For instance, a tilted beam can be produced by simultaneously exciting two different radiating modes of a meta-inspired antenna structure through probe and slot coupling techniques [18]. This configuration yields a 35° tilted beam with a realized gain of 7.9 dBi, although the operational bandwidth remains relatively narrow at 8.4%. In another implementation of this concept, an open substrate integrated waveguide (SIW) is designed to couple energy to two monopoles and a slot near its opening, resulting in an approximate 45° beam tilt

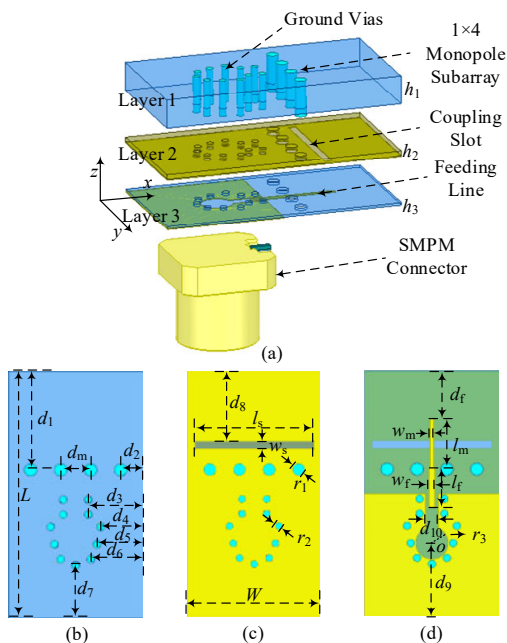

Fig. 2. Configuration of the proposed antenna: (a) Perspective view, (b) Layer 1, (c) Layer 2, (d) Layer 3.

TABLE I
DIMENSIONS OF PROPOSED TILTED BEAM ANTENNA

| Parameters | $L$ | $W$ | $d_1$ | $d_2$ | $d_3$ | $d_4$ | $d_5$ |
|---|---|---|---|---|---|---|---|
| Value (mm) | 10.0 | 5.4 | 4.0 | 0.9 | 2.2 | 1.7 | 1.85 |
| Parameters | $d_6$ | $d_7$ | $d_8$ | $d_9$ | $d_{10}$ | $r_1$ | $r_2$ |
| Value (mm) | 2.1 | 2.15 | 2.85 | 3.0 | 0.5 | 0.25 | 0.15 |
| Parameters | $r_3$ | $d_m$ | $l_s$ | $w_s$ | $l_f$ | $w_f$ | $l_m$ |
| Value (mm) | 0.65 | 1.2 | 4.8 | 0.32 | 1.6 | 0.24 | 1.98 |
| Parameters | $w_m$ | $d_f$ | $h_1$ | $h_2$ | $h_3$ | | |
| Value (mm) | 0.16 | 1.97 | 1.2 | 0.2 | 0.1 | | |

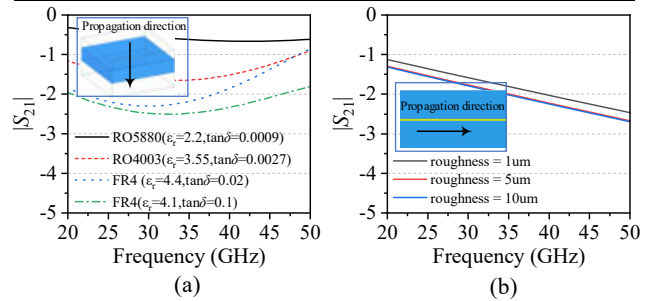

Fig. 3. Dielectric loss and conductor loss simulation results. (a) electromagnetic wave transmission loss in different substrate (same thickness of 1.2 mm). (b) Microstrip line transmission loss in FR4 substrates of different roughnesses (same thickness of 0.1mm and length of 10 mm).

and a peak gain of 5.6 dBi across a wide usable bandwidth of 49.4% [4].

This communication picks up on the underlying multi-radiator concept and addresses the drawbacks of existing beam tilting antennas [4], [7]-[18] in performance and practical implementation: 1) High cost – Existing designs are mainly implemented using printed circuit boards (PCBs) of low-loss substrates, e.g., Rogers RO4003C, instead of the low-cost but higher-loss FR4 substrate. Swapping the substrates and retuning is feasible in some cases, but even so they may by design suffer from substantially lower radiation efficiency. For example, if FR4 is used in the SIW-based design [4], high dielectric loss will result from channeling the guided wave from the source through the open SIW structure; 2) Limited bandwidth - The bandwidth of up to 49.4% is insufficient to cover all 5 G millimeter wave (mmWave) bands, e.g., Band n260 (37-40 GHz); 3) Limited tilt angle – The tilt angle of existing designs are typically below 45°, which may not be suitable for some use cases like the one depicted in Fig. 1(a); 4) Integration complexity – The integration of SIW in some antennas (e.g., [4]) with the RF circuit may require a potentially lossy transition; 5) Antenna profile – In some designs, the profile is high due to the operating mechanism. For example, in metamaterial-based designs, the spacing between the source and the metamaterial needs to meet certain requirements.

In this communication, to address the above challenges, we designed a novel low-cost wideband tilted beam antenna for mmWave vehicle communications using a step-by-step procedure. First, to achieve high radiation efficiency with low-cost lossy FR4 substrate, slot antenna is selected. Very short microstrip lines are then used to feed power in the middle of the slot to suppress the generation of even-order (i.e., non-broadside) modes in the slot. A four-element monopole array is loaded to one side of the slot, and the energy can be coupled to the monopoles through the radiation slot, so that the monopoles can generate vertical current. The horizontal electric field of the slot and the vertical electric field of the monopole array superimpose in the far-field to produce a tilted beam that is stable over a wide band, due to inherent stability of monopole and slot patterns. Moreover, to widen the impedance bandwidth, additional resonant modes are induced in both the circuit and antenna structures using the feed line, feed cavity and coupled monopole arrays. The ease of fabrication and low manufacturing cost of FR4 based design, wide impedance bandwidth through the merging of multiple modes, and stable beam pattern make this antenna ideal for vehicle wireless communications.

The rest of the paper is organized as follows: Section II presents the structure and operating principle of the proposed tilted beam antenna, whereas Section III provides the measured results and comparison with state-of-the-art designs. Conclusions are drawn in Section IV.

## II. ANTENNA STRUCTURE AND DESIGN PROCEDURE

### A. Antenna Structure

The proposed tilted-beam antenna is constructed using FR4 substrates [19] measuring 10 mm × 5.4 mm ($\varepsilon_r$ = 4.4, tan $\delta$ = 0.02 at 10 GHz) on the top and bottom layers (Layers 1 and 3), as illustrated in Fig. 2. A TU768 prepreg, a variant of FR4 material with $\varepsilon_r$ = 4.3 and tan $\delta$ = 0.023 at 10 GHz, serves as the bonding medium (Layer 2) between the substrates. Each copper layer in Layers 2 and 3 has a thickness of 0.035 mm. To minimize transmission loss, a very short microstrip line is used for signal feeding. In operation, part of the energy from the feed line radiates through a narrow slot, while the remainder is coupled to and emitted by four adjacent vertical grounded vias forming a monopole array. An SMPM connector is employed as the feed port. To ensure proper grounding and suppress energy leakage, a ring of via holes is implemented around the feed cavity near the connector. Due to fabrication constraints, these vias extend into Layers 1 and 2; however, the extension does not serve any electrical function. The detailed dimensions of the proposed antenna are listed in Table I.

### B. Design Considerations Using Lossy FR4 Substrates

To achieve a low-cost design, this work focused on using FR4 substrates. FR4 is commonly used to design sub-6 GHz antennas (e.g., smartphone antennas) since its higher loss compared to more exotic high-frequency substrates (e.g., Rogers RO4003) is acceptable due to its cost advantage. However, even higher losses in mmWave bands have so far limited FR4's use in mmWave antennas in the literature. Therefore, the first design step is to analyze antenna losses in FR4 substrates and mitigate them with explicit design choices.

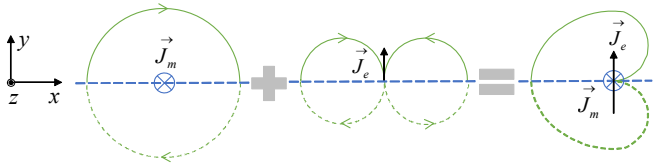

Fig. 4. Far-field synthesis of broadside and monopole-like patterns. A ground plane (– –) is assumed and only the upper half plane is considered [22].

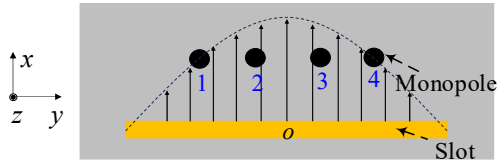

Fig. 5. Electric field distribution diagram of the slot.

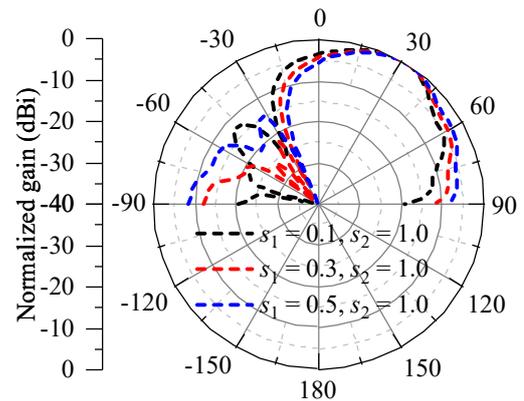

Fig. 6. Simulated synthetic pattern for different excitation amplitudes $s_1$, $s_2$.

The total (insertion) loss of the antenna can be expressed by [20]

$$\alpha_T = \alpha_C + \alpha_D + \alpha_R + \alpha_L, \quad (1)$$

where $\alpha_C$ is the conductor loss, $\alpha_D$ the dielectric loss, $\alpha_R$ the radiation loss (in the feed line) and $\alpha_L$ the leakage loss. $\alpha_L$ is typically not significant in PCBs for RF low-power applications due to their high resistivity. As for $\alpha_R$, it increases with frequency and thickness but decreases with increasing dielectric constant. Hence, it is beneficial to use a thin FR4 substrate in the transmission line, so that $\alpha_R$ can be ignored compared with $\alpha_C$ and $\alpha_D$ [20]. On this basis, using bare copper microstrip line feed is an ideal choice, and fabrication errors will lead to a lower loss [20]. However, $\alpha_C$ is heavily impacted by the feed line's surface roughness. $\alpha_D$ is the dominant loss mechanism in FR4, and it can be mitigated by limiting the amount of dielectric material in the antenna near fields, making thin planar antennas (e.g., printed dipole, slot, patch) a good choice. Moreover, to apply the microstrip line feed as well as to enable wideband property and coupling to a second radiator for beam tilting, slot antenna is chosen.

The impact of FR4 dielectric loss and feed line surface roughness on antenna gain was analyzed from 20 GHz to 50 GHz with the finite-element solver of 2020.R1 ANSYS HFSS software, based on some reference parameters [22]. The propagation losses of four substrates are compared in Fig. 3(a). It can be seen that even when the loss tangent is as high as 0.1, the maximum loss of FR4 is less than 1 dB worse than that of RO4003. As for surface roughness, the transmission loss of two kinds of microstrip lines and the ground under different roughnesses were simulated on an FR4 substrate ($\varepsilon_r$ = 4.4 and tan $\delta$ = 0.02) using the Huray model in ANSYS HFSS and plotted in Fig. 3(b). The simulated results show that when the microstrip line is relatively short (10 mm), the roughness of the surface copper has little impact on the insertion loss. The difference between the insertion loss of 1$\mu$m roughness and that of 10$\mu$m roughness is less than 0.5 dB. For comparison, this difference is less than 0.2 dB for RO4003 substrate ($\varepsilon_r$ = 3.55 and tan $\delta$ = 0.0027). Therefore, with suitable design choices (e.g., substrate thickness of up to 1.2 mm), FR4 does not incur more than 1.5 dB of extra losses in $\alpha_C$ and $\alpha_D$, as compared with high-frequency substrates, which may be acceptable as a cost-performance tradeoff.

*C. Tilted Beam Design and Beam Stability*

Following the choice of slot antenna with microstrip line feed on a thin FR4 substrate, the next step is to tilt the antenna pattern. The beam-tilting operation of the proposed antenna is inspired by the recent design [4] that utilizes the idea of complementary sources [22]. As shown in Fig. 4, due to the current directions, combining a broadside pattern from a magnetic current source $\vec{J}_m$ (e.g., the slot) with a monopole-like pattern from an electric current source $\vec{J}_e$ (i.e., second radiator) produces an equivalent tilted beam pattern [4]. Hence, (slot-fed) monopole is a natural choice for the second radiator.

Figure 5 presents the electric field distribution of the *y*-oriented half-wave slot (of length $L$) for the fundamental mode, given by [23]

$$E_s(z) = E_0 \cos(\pi y / L), -L/2 \le y \le L/2, \quad (2)$$

where $E_0$ is the maximum amplitude of the electric field. Therefore, to excite the monopole by coupling feed, it should be placed near the center of the slot. To enhance the excitation of the monopole pattern and increase the gain, two monopoles (with spacing $d$) were utilized. Moreover, two more monopoles were added to either side of the two-monopole array, where the fields are weaker (see Fig. 5). As will be elaborated in the next section, the two outer monopoles mainly serve to improve impedance matching and hence the bandwidth.

Furthermore, to improve the sidelobe levels of the tilted beam, the relative level of excitation between the slot and monopole array needs to be optimized. This effect can be demonstrated analytically with closed-form antenna patterns. Specifically, the normalized pattern of a monopole in the $\theta$-cut can be expressed as [23]

$$E_m = (E_{\theta 1} + E_{\theta 2}) / \max(E_{\theta 1} + E_{\theta 2}), \quad (3)$$

where

$$E_{\theta 1} = jkZ_0 e^{-jkr_0} / (4\pi r_0) \int_0^H I(z) e^{-jkz\cos\theta} \sin\theta dz, \quad (4)$$

$$E_{\theta 2} = kZ_0 \cos\theta e^{-jkr_0} / (2r_0) \int_0^a J(\rho) J_1(k\rho\sin\theta)\rho d\rho, \quad (5)$$

where $k$ is the wave number, $J(\rho)$ is the current expression (11) in [24], $H$ the equivalent physical length of the monopole antenna and $a$ is the radius of the ground area. $Z_0$ is the wave impedance of free space, $r_0$ is the distance to the observation point in the far-field, and $\theta$ is the polar angle relative to the axis perpendicular to the structure. $J_1(\cdot)$ is the first-order Bessel function of the first kind. Focusing only on the two middle monopoles that dominantly contribute to the radiation, the excitation amplitudes are identical and denoted by $s_2$. The array factor is then [23],

$$AF = \left| \frac{\sin(\pi N_x d_x \sin\theta/\lambda)}{N_x \sin(\pi d_x \sin\theta/\lambda)} \right| \cdot \left| \frac{\sin(\pi N_y d_y \sin\phi/\lambda)}{N_y \sin(\pi d_y \sin\phi/\lambda)} \right| \quad (6)$$

where $N_{x/y}$ and $d_{x/y}$ are the number of elements and spacing along the *x/y* directions, respectively. Considering the plane $\phi = 0°$ and $N_x = 1$, $N_y = 2$, $d_y = d$, then $AF = 1$. On the other hand, according to cavity theory [23], the normalized pattern (with $\phi = 0°$) of the slot antenna (magnetic source) for the fundamental mode is

$$E_s(\theta) = \sin\left(\frac{\pi}{2}\sin\theta\right). \tag{7}$$

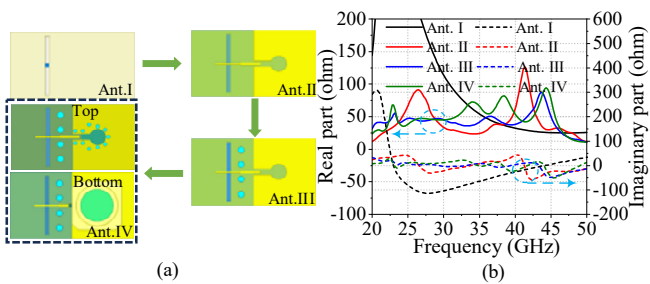

Fig. 7. Bandwidth enhancement procedure: (a) Evolution of the proposed antenna in four steps and (b) corresponding input impedances.

Since the distance between the slot and the monopole array is very small relative to the far-field observation point, they are considered to have the same phase center. The total far-field pattern is then

$$E_{total}(\theta) \approx s_1 \cdot E_s(\theta) + s_2 \cdot E_m^{23}(\theta), \tag{8}$$

where $s_1$ is the excitation amplitude of the slot. Using (6), the impact on the radiation patterns for different excitation amplitudes is evaluated. The slot's first and third modes have similar patterns in the $\phi = 0°$ direction but different beam widths. Therefore, only the results for the first mode are shown in Fig. 6. It can be seen that when the excitation amplitude ratio of the two radiators is 0.3, the overall pattern of the antenna has a lower sidelobe. Therefore, the sidelobe can be reduced by adjusting the distance (and hence coupling) between the slot and the monopole.

In this analytical study, the monopole is assumed to be on an infinite ground. In practice, due to the influence of limited ground, the maximum gain direction will be about $\theta = 60°$. At the same time, since the equivalent monopole length is between quarter- and half-wavelength over the operating frequency band, although the gain changes, the far-field radiation pattern remains monopole-like, so it has little impact on the synthesized pattern. Moreover, it can be seen that even when there is a difference in the excitation amplitudes between the two radiators, the direction of beam deflection is almost unchanged. This shows that even as the operating frequency moves away from the center frequency, the beam direction is relative stable, despite changes in the relative excitation amplitudes of the slot and monopole radiators. This in turn indicates the inherent wideband property of the tilted beam. Similarly, the quasi-second-order mode of the slot and the monopole array also produce tilted beam patterns.

*D. Impedance Bandwidth Enhancement*

The input impedance plot of Fig. 7 illustrates the procedure to enhance bandwidth. Ant. I, as a basic slot antenna, has a resonance near 22 GHz, but the imaginary part shows high inductance and the real part impedance is also high, resulting in poor matching. Ant. II uses a microstrip line with open terminals to feed the basic slot antenna (i.e., Ant. I). Due to the feed structure, the slot resonance increases to around 26 GHz, and the real part of the impedance is greatly reduced. At the same time, due to the microstrip line being short, a second resonance occurs at 42 GHz. This resonant frequency from the short microstrip line can be approximated as half-wave transmission line mode, i.e.,

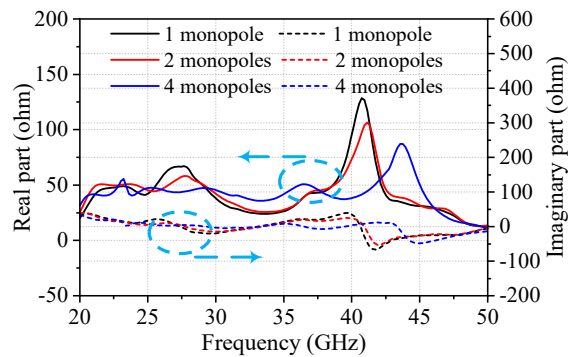

Fig. 8. Effect of the number of monopoles on antenna impedance.

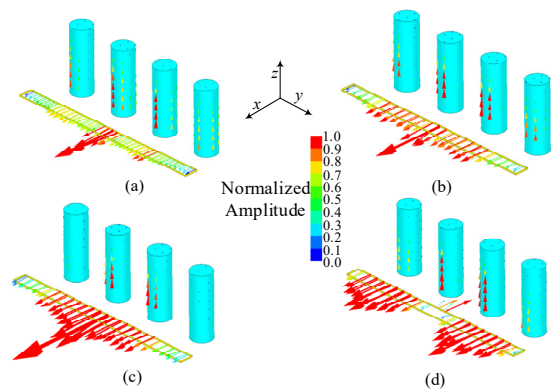

Fig. 9. Electric field along slot edge nearest to monopole array and monopole array current distribution at: (a) 26 GHz. (b) 31 GHz. (c) 36 GHz. (d) 41 GHz.

$$f_m = \frac{c}{2l_m\sqrt{\varepsilon_r}} \tag{9}$$

where $c$ is the speed of light in the vacuum and $l_m$ is the length of the open terminal for the microstrip line in Fig. 2(d). However, the impedances at the two resonances are still high, and the two resonances are far apart, making it difficult to combine them.

To further improve matching (apart from enabling monopole mode for beam tilting, as described in Section II-C), four monopoles are added to the side of the slot closer to the feeding port (Ant. III). To motivate the choice of four monopoles, the impact of the number of monopoles on the input impedance (in the setup of Ant. III) is depicted in Fig. 8. When only one monopole is placed close to the middle of the slot (i.e., one of the two middle positions), in addition to retaining Ant. II's resonances (i.e., 26 GHz and 42 GHz), a further resonance appears at 23 GHz. When only two middle monopoles are loaded, the real parts of the impedance at the resonances near 26 GHz and 42 GHz move closer to 50 ohms. Finally, when the number of loaded monopoles increases to four, except for the resonance near 23 GHz, the impedance of other resonances approaches 50 ohms. Moreover, an additional resonance point is created near 29 GHz. Hence, as the number of monopoles increases, the number of resonances increases and in general, the increased monopole loading improves the matching. This is because the interactions between the slot and the monopoles, as well as that among the monopoles, create additional resonances and transform the impedance.

In the final step, an SMPM connector is mounted on the back of the antenna (Ant. IV) to feed the microstrip line. This configuration introduces a slight impact on the impedance performance within the 30–40 GHz range. The back-mounted SMPM connector is selected instead of a Southwest Microwave edge-mount connector to keep the microstrip line as short as possible and to minimize the connector's influence on the antenna's radiation characteristics.

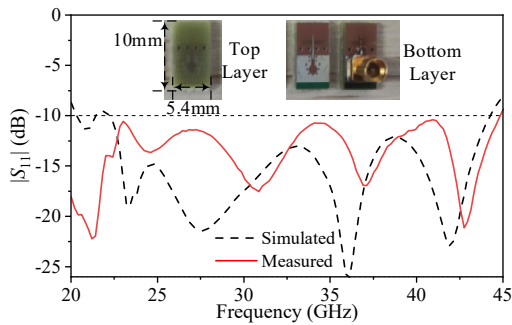

Fig. 10. Simulated and measured $|S_{11}|$ of the proposed antenna.

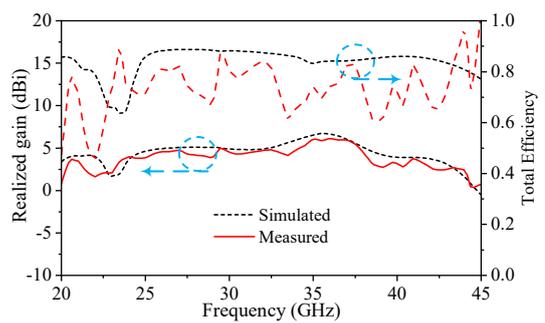

Fig. 12. Simulated and measured realized gain of the proposed antenna.

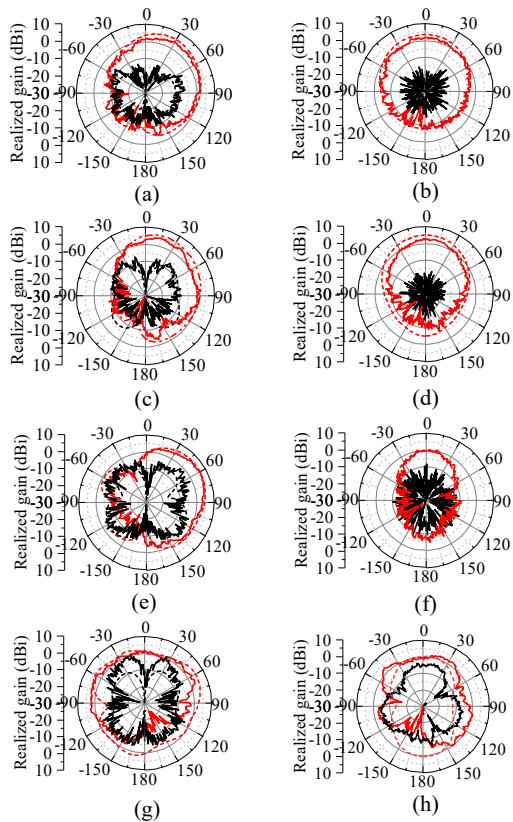

Fig. 11. Simulated and measured radiation patterns of the proposed titled beam antenna. (a) E-plane at 26 GHz. (b) H-plane at 26 GHz. (c) E-plane at 31 GHz. (d) H-plane at 31 GHz. (e) E-plane at 36 GHz. (f) H-plane at 36 GHz. (g) E-plane at 41 GHz. (h) H-plane at 41 GHz.

To further elucidate the operating principles of the proposed antenna (Ant IV), Fig. 9 presents the electric field distribution of the slot and the surface current distribution of the monopole array at its four resonant frequencies: 26 GHz, 31 GHz, 36 GHz, and 41 GHz. For clarity, only the slot and monopole arrays are illustrated. As observed, the surface current distributions on the monopole array exhibit similar patterns across all four frequencies, with the primary variation arising from the electric field distribution within the slot. At 26 GHz, the electric field is predominantly concentrated near the center of the slot. At 31 GHz and 36 GHz, while the strongest field still appears at the center, the electric field becomes more evenly distributed along the slot length compared to the 26 GHz case. At 41 GHz, the in-phase electric field is primarily concentrated near one-quarter and three-quarters of the slot length, indicating the excitation of a higher-order mode. Although there are slight differences in the current magnitudes among the monopoles, the vertical orientation of these currents is sufficient to tilt the beam generated by the horizontal currents in the slot.

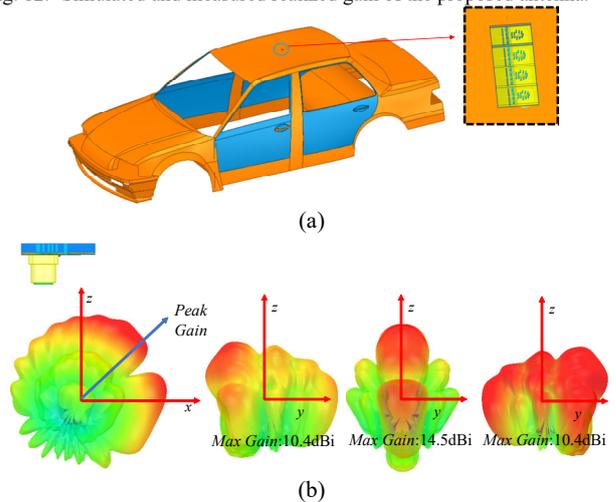

Fig. 13. Array beam scanning: (a) Simulation of a 1×4 rooftop array; (b) Result of beam scanning.

III. EXPERIMENTAL VALIDATION AND PERFORMANCE COMPARISON

For experimental validation, a prototype of the proposed antenna is fabricated with a multi-layer process and a SMPM connector is used (see Fig. 10). The prototype occupies an area of $10 \times 5.4 \times 1.5$ mm$^3$. The simulated and measured 10 dB impedance bandwidth are 63.48% (22.41-43.25 GHz) and 76.51% (20-44.78 GHz), respectively (see Fig. 12). The radiation patterns are shown in Fig. 11 for 26 GHz, 31 GHz, 36 GHz and 41 GHz. The simulated and measured cross-polarization gain is at least 20 dB and 15 dB lower than that of the co-polarization in the direction of maximum gain, respectively. The gain and efficiency results are shown in Fig. 12. The simulated realized gain varies from 1.6 dBi to 6.7 dBi. The measured peak gain is 0.5-1.3 dB less than the simulated one due to connector loss and the maximum measured gain can reach 6.1 dBi. Overall, given fabrication and tolerance issues in mmWave antenna design, the simulated and measured results are in reasonable agreement.

Table II gives performance comparisons of the proposed antenna with state-of-the-art tilted beam antennas that are representative of different beam-tilting approaches. An impedance bandwidth close to 50% can be obtained in [4], but its structure requires a certain length, resulting in a larger size. For solutions that use metasurface loading methods to generate tilted beams [7], [12], their profiles are relatively high, often greater than $0.2\lambda_0$. Although [17] can also obtain a wider bandwidth (40%) and a lower profile, its pattern is bidirectional. A gain of 7.9 dBi and a profile of $0.048\lambda_0$ have been achieved using a meta-inspired antenna to excite two radiating modes [18], but the tilt angle is limited (35°) and the bandwidth is narrow (8.4%). In contrast, the proposed antenna is not only unique in its use of low-cost FR4 substrate, explicit design considerations facilitates its competitive performance: the largest bandwidth, second largest tilt angle, as well as the highest peak gain for antennas with over 30% impedance bandwidth.

TABLE II
COMPARISON BETWEEN THE PROPOSED ANTENNA AND STATE-OF-THE-ART ANTENNAS WITH TILTED BEAMS

| Ref | $f_0$ (GHz) | BW (%) | Peak Gain (dBi) | Tilt Angle* (°) | Sizes ($\lambda_0 \times \lambda_0 \times \lambda_0$) | Cost |
|---|---|---|---|---|---|---|
| [4] | 31 | 49.4 | 5.6 | 45 | 0.64×1.70×0.16 | High (RO4003C) |
| [7] | 8 | 5.2 | 16.4 | 33 | 3.60×3.60×0.48 | High (substrates of tan $\delta$ < 0.0015) |
| [12] | 4.4 | 8.0 | 6.2 | 44 | 2.20×1.76×0.4 | High (RO4003C) |
| [16] | 11.2 | 10.7 | 6 | 33 | 0.24×0.24×0.45 | High (resonator of tan $\delta$ < 0.0009) |
| [17]† | 4.75 | 32 | 5 | 40 | 1.36×1.16×0.025 | High (RO5880) |
| [18] | 5.1 | 8.4 | 7.9 | 35 | 1.96×1.96×0.048 | High (F4B with tan $\delta$ = 0.002) |
| This Work | 32.4 | 76.5 | 6.1 | 46 | 0.59×1.10×0.17 | Low (FR4) |

\* The tilt angle is taken at the center frequency, where specified.
† The ground plane size of $1.73\lambda_0 \times 1.73\lambda_0$ not included.

Fig. 13 illustrates the simulated beam scanning results for a 1×4 sub-array mounted on the rooftop of a vehicle. In Fig. 13(b), the direction of maximum radiation gain is observed to tilt toward the roadside, facilitating effective communication with RCD deployed along the road. Beam patterns are presented for steering angles of –45°, 0°, and +45°. The array achieves a peak gain of 14.5 dBi at boresight (0°), while gain degradation of approximately 4 dB is noted at ±45°, consistent with the expected beam squint and reduced array aperture projection. Notably, metallic reflections from the vehicle rooftop lead to multipath interference, superimposing secondary lobes onto the main beam. This results in observable fluctuations in the main lobe gain, potentially affecting link stability and reliability in dynamic environments.

IV. CONCLUSION

In this communication, a novel wideband tilted-beam antenna is proposed for mmWave vehicle communication applications. The antenna achieves beam tilting through the far-field superposition of the horizontal electric field generated by the slot and the vertical electric field radiated by the monopole array. To validate the proposed design, a prototype was fabricated. Measured results demonstrate a -10 dB impedance bandwidth of 76.5% (20–44.78 GHz) and a peak realized gain of 6.09 dBi. The antenna features wide bandwidth, a stable tilted beam, and low-cost implementation, making it a strong candidate for mmWave vehicle wireless communication and sensing applications, such as smart sensors mounted at RCD.

ACKNOWLEDGMENT

The authors want to thank Mr. B. K. Lau for the improvement of writing and Mr. Y. Shi for the help with antenna measurement.